\documentstyle[prb,aps,multicol,floats]{revtex}

\input{epsf}
\epsfverbosetrue

\begin{document}
\twocolumn[\hsize\textwidth\columnwidth\hsize\csname @twocolumnfalse\endcsname

\title{Electronic and vibrational properties of low-dimensional perovskites
    Sr$_{1-y}$La$_y$NbO$_{3.5-x}$}

\author{C. A. Kuntscher}
\address{Forschungszentrum Karlsruhe, Institut f\"ur Festk\"orperphysik,
D-76021 Karlsruhe, Germany  \\
and 1. Physikalisches Institut, Universit\"at Stuttgart, Pfaffenwaldring 57,
D-70550 Stuttgart, Germany}

\author{S. Schuppler}
\address{Forschungszentrum Karlsruhe, Institut f\"ur Festk\"orperphysik,
D-76021 Karlsruhe, Germany}

\author{P. Haas, B. Gorshunov,$^*$ and M. Dressel}
\address{1. Physikalisches Institut, Universit\"at Stuttgart, Pfaffenwaldring 57,
D-70550 Stuttgart, Germany}

\author{M. Grioni}
\address{Institut de Physique Appliqu\'ee, Ecole Polytechnique
F\'ed\'erale, CH-1015 Lausanne, Switzerland}

\author{F. Lichtenberg}
\address{Experimentalphysik VI, Institut f\"ur Physik, EKM,
Universit\"at Augsburg, Universit\"atsstr. 1, D-86135 Augsburg, Germany}

\date{\today}
\maketitle

\begin{abstract}
By angle-resolved photoemission spectroscopy and polarization-dependent infrared
reflectivity measurements the electronic and vibrational properties
of low-dimensional perovskites Sr$_{1-y}$La$_y$NbO$_{3.5-x}$
were studied along different crystal directions. The electronic behavior
strongly depends on the oxygen and La content, including quasi-one-dimensional
metallic and ferroelectric insulating behavior.
An extremely small energy gap at the Fermi level is found for SrNbO$_{3.41}$
and SrNbO$_{3.45}$ along the conducting direction at low temperature,
suggestive for a Peierls-type instability. Despite the similar nominal carrier
density, for Sr$_{0.8}$La$_{0.2}$NbO$_{3.50}$ the quasi-one-dimensional metallic
character is suppressed and no gap opening is observed, which can be explained
by differences in the crystal structure. Electron-phonon interaction appears
to play an important role in this series of compounds.
\end{abstract}

]

\section{Introduction}
\begin{figure}
\hspace*{0mm} \leavevmode \epsfxsize=85mm \epsfbox{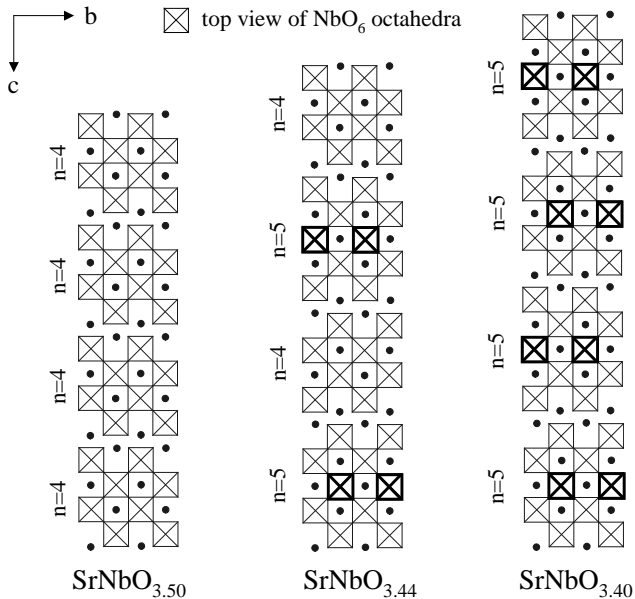}
\caption{Schematic drawing of the idealized, i.e., non-distorted, crystal
structures of three compounds of the series Sr$_n$Nb$_n$O$_{3n+2}$.
The NbO$_6$ units in the middle of the slabs, which presumably are the
conducting channels of the lattice, are marked by thick lines.}
\label{crystal0}
\end{figure}

The observation of quasi-one-dimensional (quasi-1D) metallic behavior
of some members of the perovskite-related series Sr$_{1-y}$La$_y$NbO$_{3.5-x}$, with
0$\leq$$x$$\leq$0.1, raised interest recently.\cite{Kuntscher00,Kuntscher02}
It comprises compounds with a wide range of electronic properties,\cite{Lichtenberg01}
including a ferroelectric insulator, semiconducting compounds, and quasi-1D metals.
They have an orthorhombic, perovskite-related crystal structure with
lattice parameters $a$$\approx$3.9 \AA, $b$$\approx$5.5 \AA, and $c$
depending on the oxygen content,\cite{Lichtenberg01,Abrahams98,Williams93}
and belong to the homologous series of the type
Sr$_n$Nb$_n$O$_{3n+2}$. The main building blocks of the
structure are NbO$_6$ octahedra, which are grouped into slabs extending
along the ($a$,$b$) plane. The width of the slabs along $c$ is determined
by the oxygen content and directly given by the parameter $n$:
While for quasi-1D metallic SrNbO$_{3.40}$ ($n$=5) the slabs are five octahedra
wide, ferroelectric, insulating SrNbO$_{3.50}$ ($n$=4) consists of four
octahedra wide slabs. For defined oxygen stoichiometries in the range
0$\leq$$x$$\leq$0.1 the structure exhibits a well-ordered stacking
sequence of the ($n$=5)- and ($n$=4)-subunits, listed in Table \ref{stacking}
for the investigated compounds.\cite{Lichtenberg01,Williams93,Bednorz97}
A schematic drawing of the idealized crystal structure of three
representative SrNbO$_{3.50-x}$ compounds is shown in Fig.\ \ref{crystal0}.
The real crystal structure exhibits a considerable distortion of the
NbO$_6$ octahedra, with different magnitude depending on the position
within the slab. Along the $a$ axis the octahedra are connected
continuously via the apical oxygen sites forming 1D chains,
as is illustrated in Ref.\ \onlinecite{Kuntscher02}.
The oxygen content not only determines the crystal structure but also the
number of Nb$4d$ conduction electrons. A second doping channel which,
however, does not affect the width and stacking sequence of the slabs is
the substitution of divalent Sr by a higher-valent element like La.
The nominal numbers of Nb$4d$ conduction electrons per formula unit
for the present niobates, derived within an ionic picture, are listed
in Table \ref{stacking}.

\begin{table}
\caption{Studied niobates of the type Sr$_n$Nb$_n$O$_{3n+2}$.
For each composition the stacking sequence of the (n=5)- and (n=4)-subunits
per unit cell and the nominal number $z$ of Nb$4d$ conduction electrons
per formula unit are given.}
\label{stacking}
\centering
\begin{tabular}{ccc}
                                 & stacking & $z$ \\
                                 & sequence  &     \\
\hline
SrNbO$_{3.41}$                   & -5-5-5-5- & 0.18 \\
SrNbO$_{3.45}$                   & -4-5-4-5- & 0.1 \\
SrNbO$_{3.50}$                   & -4-4-4-4- & 0 \\
Sr$_{0.8}$La$_{0.2}$NbO$_{3.50}$  & -4-4-4-4- & 0.2 \\
Sr$_{0.7}$Ba$_{0.1}$La$_{0.2}$NbO$_{3.50}$ & -4-4-4-4- & 0.2 \\
Sr$_{0.6}$Ca$_{0.2}$La$_{0.2}$NbO$_{3.50}$ & -4-4-4-4- & 0.2
\end{tabular}
\end{table}

According to dc resistivity measurements, the electronic properties of
SrNbO$_{3.41}$ and Sr$_{0.8}$La$_{0.2}$NbO$_{3.50}$ are highly anisotropic,
with lowest, metal-like resistivity values along the $a$ direction.\cite{Lichtenberg01}
The temperature dependence of the resistivity along the $a$ axis, $\rho_a$,
consists of several regimes; taking SrNbO$_{3.41}$ as an example, one observes
the following behavior:
Starting from the lowest temperature, the resistivity decreases strongly,
showing a minimum around 50 K and increases between 50 and $\approx$130 K.
Above $\approx$130 K $\rho_a$ decreases slightly with increasing temperature.
The resistivity behavior below 50 K can be described by an activated behavior
$\rho$$\propto$$\exp[E_a/(k_BT)]$ with activation energies
$E_a$=2-3 meV.\cite{Kuntscher02,Lichtenberg01} An explanation for the temperature
behavior of $\rho_a$ above 50 K is so far missing. A similar temperature 
dependence is observed for Sr$_{0.8}$La$_{0.2}$NbO$_{3.50}$.

On the other hand, SrNbO$_{3.50}$ ($n$=4) is a band insulator.
It is ferroelectric below $T_c$=1613 K with a spontaneous polarization along
the $b$ axis.\cite{Nanamatsu71,Nanamatsu75} The compound undergoes two further
phase transitions, namely a second-order phase transition to an incommensurate
phase at 488 K and another ferroelectric transition at 117 K with an
additional component of the spontaneous polarization appearing along the
$c$ axis.\cite{Ohi79}

The magnetic properties of the investigated niobium oxide compounds
were described in Ref.\ \onlinecite{Lichtenberg01}. Except SrNbO$_{3.50}$,
all compounds show a similar behavior: With increasing temperature, between
4 and 50 K the magnetic susceptibility decreases steeply due to paramagnetic
impurities, followed by an increase between 50 and 400 K. This increase is
not yet understood; explanations in terms
of thermally activated charge carriers excited across an energy gap or
the Bonner-Fisher model for a 1D antiferromagnetic spin-1/2 Heisenberg
spin chain were proposed.\cite{Weber01}

The conduction mechanism of the compounds SrNbO$_{3.41}$,
SrNbO$_{3.45}$, and Sr$_{0.8}$La$_{0.2}$NbO$_{3.50}$ was studied by dielectric
measurements in the radio-frequency range along the $c$ direction.\cite{Bobnar02}
They suggest a hopping transport of charge carriers with polaronic character
along $c$.
The results from EPR studies on SrNbO$_{3.41}$\cite{Weber01} with the
magnetic field applied parallel to the $b$ and $c$ axis are in
agreement with a variable range hopping in 1D. Furthermore, $^{93}$Nb NMR
measurements indicate a fluctuating quadrupole field and seem to resemble
some of the characteristics of the prototypical charge-density-wave (CDW)
compound NbSe$_3$.\cite{Weber01}

The most puzzling finding for the series \linebreak Sr$_{1-y}$La$_y$NbO$_{3.50-x}$
is an extremely small energy gap at the Fermi energy for SrNbO$_{3.41}$ along the
chain direction.\cite{Kuntscher02} In this manuscript we will show that
signatures of a very small energy gap are also present for the compound
SrNbO$_{3.45}$. In order to understand the observed quasi-1D metallic character
and the unusual properties at low temperature, we
give an overview of the electronic and vibrational properties of various
compounds of the series Sr$_{1-y}$La$_y$NbO$_{3.5-x}$, as obtained by 
optical spectroscopy in the infrared frequency range and
angle-resolved photoemission spectroscopy (ARPES). We will discuss our findings
for the various compounds with respect to differences in the crystal structure.
Based on the comparison of our results from optics and ARPES to the dc 
resistivity data, we propose a possible picture for the transport mechanism 
and discuss the role of electron-phonon coupling in these niobate compounds.

The manuscript is organized as follows:
The experimental details are given in Sec.\ II, followed by the
presentation of the experimental results in Sec.\ III. We discuss the
electronic and vibrational properties of the series Sr$_{1-y}$La$_y$NbO$_{3.5-x}$
in Sec.\ IV and conclude with a summary of the findings in Sec.\ V.

\section{Experiment}
\label{sectionexperiment}
The investigated SrNbO$_{3.41}$, SrNbO$_{3.45}$, SrNbO$_{3.50}$, and
(Sr,Ba,Ca)$_{0.8}$La$_{0.2}$NbO$_{3.50}$
single crystals with a typical size of 3$\times$2$\times$0.2~mm$^3$ were
grown by a floating zone melting process, and their oxygen content was
determined by thermogravimetric analysis.\cite{Lichtenberg01}
ARPES data were recorded with a Scienta ESCA-300 analyzer. The He I$\alpha$
(21.22 eV) line of a standard gas discharge
lamp was used. With a fixed analyzer setup the
sample was rotated around one axis; the rotation angle $\theta$
determines the emission direction of the photoelectrons relative to
the surface normal and thus the component $k$ of the photoelectron
momentum along the studied direction according to
$k$=$\sqrt{2mE_{kin}} \sin\theta/\hbar$.
The crystals, which had been oriented to $\pm$1$^{\circ}$ by Laue
diffraction before introducing them into the UHV system, were cleaved
at a base pressure of 1$\times$10$^{-10}$ mbar and cooled down.
ARPES spectra with energy resolution $\Delta$$E$=30 meV
were recorded at temperature $T$=75 K. The sample surface quality
was checked periodically by comparison of valence band features.
The angular resolution of the ARPES spectra amounts to
$\Delta\theta$=$\pm$0.5$^{\circ}$ corresponding to $\Delta$$k$=0.04~\AA$^{-1}$.
The Fermi energy $E_F$ was determined to 1 meV accuracy from the Fermi
cutoff of a freshly evaporated gold film recorded immediately afterwards
at the same experimental conditions.

For the reflectivity measurements the samples were carefully polished  on 
a polishing disc using diamond paste with a grain size of 1 $\mu$m to obtain
flat and shiny surfaces.
Near-normal incidence reflection measurements with the electric field {\textbf E}
of the incident light parallel to the $a$ and $b$ axes were performed for frequencies
20-10\,000 cm$^{-1}$ using a Fourier-transform spectrometer Bruker IFS 113v.
The energy resolution was set to 1 cm$^{-1}$ in the far-infrared
range  and to 2-4 cm$^{-1}$ in the mid-infrared (MIR) and near-infrared range.
For SrNbO$_{3.41}$, SrNbO$_{3.45}$, and Sr$_{0.8}$\-La$_{0.2}$\-NbO$_{3.50}$
the low frequency reflectivity (6-30 cm$^{-1}$) was studied with a quasi-optical
submillimeter (submm) spectrometer\cite{Kozlov98} using backward-wave oscillators
which produce monochromatic, polarized radiation continuously tunable in frequency.
For SrNbO$_{3.41}$ the reflectivity spectra were extended to
34\,000 cm$^{-1}$ by utilizing a Bruker IFS 66v/S spectrometer.
The use of optical cryostates enabled temperature-dependent measurements
between 5 and 300 K. To obtain absolute reflectivities the spectra were
divided by the reflectivity spectra of an Al mirror.

\begin{figure}
\hspace*{5mm} \leavevmode \epsfxsize=65mm \epsfbox{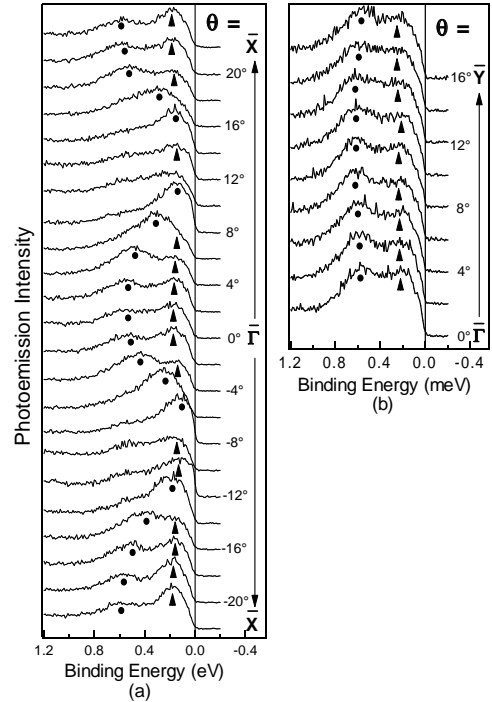}
\caption{ARPES spectra of SrNbO$_{3.41}$, normalized to the same total intensity,
along (a) $\bar\Gamma$-$\bar X$ and (b) $\bar\Gamma$-$\bar Y$ at 75 K with
$\Delta$$E$=30 meV.}
\label{srnbo341_1}
\end{figure}

To obtain the optical conductivity $\sigma_1(\nu)$ by Kramers-Kronig analysis the
reflectivity spectra $R(\nu)$ of SrNbO$_{3.41}$ and SrNbO$_{3.45}$ along
the highly conducting direction were extrapolated to zero frequency according
to the Hagen-Rubens relation.\cite{Dressel01} A constant extrapolation
to $\nu$=0 was chosen for the perpendicular direction.
For insulating SrNbO$_{3.50}$ the value $R(0)$ for the constant low-frequency
extrapolation was calculated from the dielectric constant $\epsilon_1$ at 1 kHz
($\epsilon_{1,a}$=75, $\epsilon_{1,b}$=43) \cite{Nanamatsu75}
according to
$R(0)=[(1-\sqrt{\epsilon_1})/(1+\sqrt{\epsilon_1})]^2$.
For the Sr$_{0.2}$La$_{0.8}$NbO$_{3.50}$ spectra either a Hagen-Rubens
or a constant extrapolation was used, depending on the low-frequency
behavior of the measured reflectivity at different temperatures. For the
high-frequency extrapolation of all spectra an $\nu^{-4}$ decay was
assumed.

In the case of SrNbO$_{3.45}$ the reflectivity data were supplemented
by transmission measurements in the frequency range 10-20 cm$^{-1}$
on a single crystal of thickness $\approx$20 $\mu$m using a
Mach-Zehnder interferometer arrangement. From the independent
determination of the transmission coefficient and the phase shift we
were able to directly calculate the complex dielectric function
$\epsilon(\nu)$ by utilizing the Fresnel equations.\cite{Kozlov98}

\section{Results}
\label{sectionresults}
\subsection{SrNbO$_{3.41}$: ARPES, optics}  \label{subsectionsrnbo341}
The $k$-dependent near-$E_F$ electronic structure, as studied by ARPES, reflects
the strongly anisotropic character of SrNbO$_{3.41}$ (see Fig.\ \ref{srnbo341_1}):
One observes two bands near $E_F$, with $\approx$600 and $\approx$200 meV
binding energy (BE) at the $\bar\Gamma$ point. Along $\bar\Gamma$-$\bar X$
(Fig.\ \ref{srnbo341_1}a), i.e., along the chains,
the band at lower BE shows no dispersion, whereas the second band disperses
strongly towards $E_F$, gains intensity at around $|\theta|$=8$^{\circ}$,
and loses intensity between $|\theta|$=8$^{\circ}$ and
$|\theta|$=10$^{\circ}$ suggestive of a Fermi surface (FS) crossing.
At $|\theta|$$\pm$14$^{\circ}$ the peak reappears and disperses away
from $E_F$ for higher $|\theta|$.
The reappearance of the peak can be explained in terms of a 2$\times$1
superstructure which was also observed by low energy electron diffraction
\cite{Kuntscher00,Kuntscher00a} and neutron scattering,\cite{Pintschovius}
and seems to be a characteristic property of all members of the
Sr$_{1-y}$La$_y$NbO$_{3.5-x}$ series. The parabolic dispersion of the band
around $\bar\Gamma$ and $\bar X$ gives an effective electron mass of
m$^*$/m$_0$=0.71. Along the perpendicular direction $\bar\Gamma$-$\bar Y$
(Fig.\ \ref{srnbo341_1}b) both bands hardly show any dispersion.

\begin{figure}
\hspace*{5mm} \leavevmode \epsfxsize=65mm \epsfbox{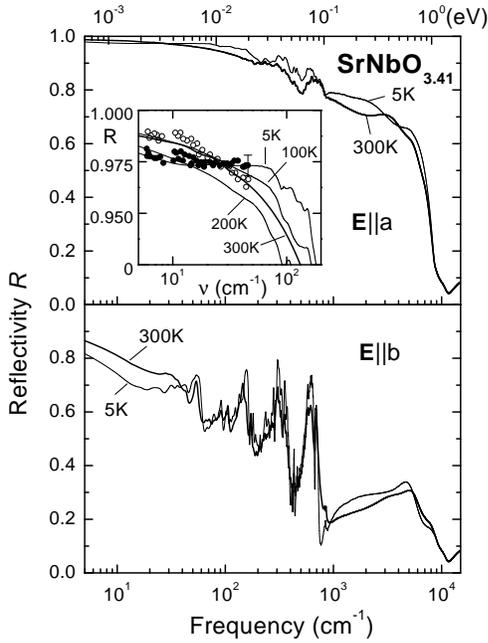}
\caption{Reflectivity $R$ of SrNbO$_{3.41}$ at different temperatures
for {\bf E}$\parallel$$a$ and {\bf E}$\parallel$$b$, respectively.
Inset: Low-frequency reflectivity for \textbf{E}$||$$a$; open (closed)
circles: submm reflectivity data at 300 K (5 K).}
\label{refl_34}
\end{figure}

\begin{figure}
\hspace*{5mm} \leavevmode \epsfxsize=65mm \epsfbox{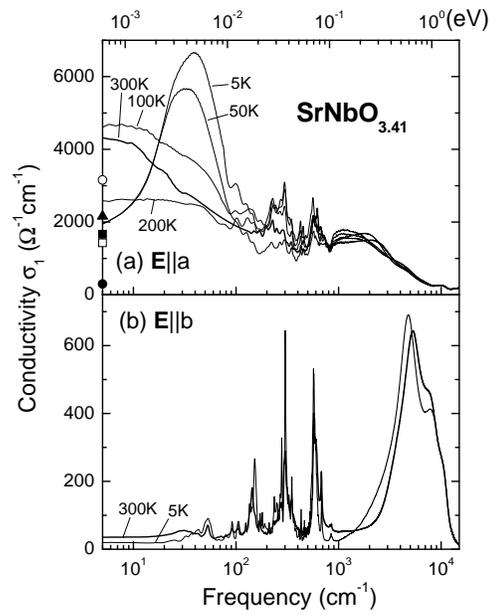}
\caption{Optical conductivity $\sigma_1$ of SrNbO$_{3.41}$
at different temperatures for {\bf E}$\parallel$$a$ and {\bf E}$\parallel$$b$,
respectively. For {\bf E}$\parallel$$a$ the low-temperature
peak around 40 cm$^{-1}$ indicates excitations across a gap
2$\Delta$$\approx$5 meV. Also shown are the dc conductivity values for the 
$a$ axis at 300 K (filled triangle), 200 K (open square), 100 K (filled square),
50 K (open circle), and 5 K (filled circle).}
\label{sig_34}
\end{figure}

The anisotropic character is confirmed by the
strongly polarization-dependent reflectivity $R(\nu)$
shown in Fig.~\ref{refl_34}. It exhibits a plasma edge for
{\bf E}$\parallel$$a$ and high values ($\approx$1) at low frequencies
indicating metallic behavior. In contrast, for {\bf E}$\parallel$$b$ the
considerably lower reflectivity demonstrates the insulating character
perpendicular to the chains. The corresponding optical conductivity $\sigma_1$
(Fig.\ \ref{sig_34}) for {\bf E}$\parallel$$a$ consists of a Drude-like
component followed by phonon modes and a relatively broad MIR band
centered around 1500 cm$^{-1}$.
A rich phonon spectrum is observed along both directions;
for {\bf E}$\parallel$$b$ 76 lines can be resolved at RT, with the strongest
modes at 144, 296, 304, 561, 572, 577, 583, 595, 603, 612, and 680 cm$^{-1}$.
The strong double-peak structure in the {\bf E}$\parallel$$b$ spectrum centered
around 5000 cm$^{-1}$ can be related to interband transitions.

\begin{figure}
\hspace*{5mm} \leavevmode \epsfxsize=65mm \epsfbox{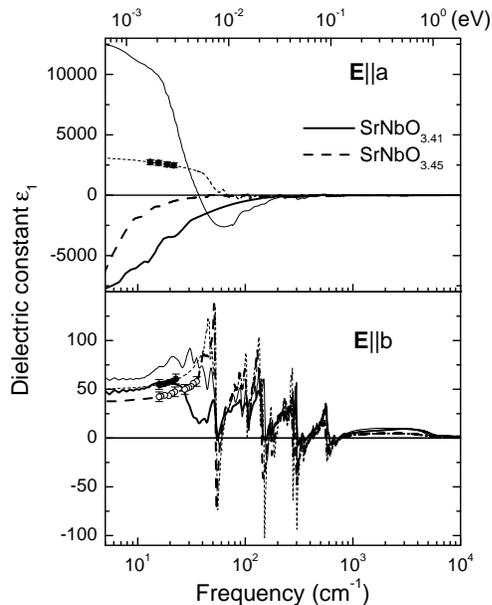}
\caption{Dielectric constant $\epsilon_1$ of SrNbO$_{3.41}$ and SrNbO$_{3.45}$
at 300 K (thick lines) and 5 K (thin lines) for {\bf E}$\parallel$$a$ and
{\bf E}$\parallel$$b$, respectively; open (closed) circles: data points at
300 K (5 K) directly determined by transmission measurements on a thin
platelet.}
\label{eps1}
\end{figure}

During cooling down considerable changes can be seen in the electronic
and vibrational part of the spectra: For {\bf E}$\parallel$$b$ at least six new
modes appear with decreasing temperature. Along the perpendicular direction,
{\bf E}$\parallel$$a$, a strong peak appears around 40 cm$^{-1}$
in the optical conductivity spectrum (see Fig.\ \ref{sig_34}), which
starts building up below 100 K. With decreasing temperature the peak
shifts slightly to higher frequencies and becomes stronger. As discussed
in Ref.\ \onlinecite{Kuntscher02} this feature can be ascribed to
single-particle excitations across a gap in the
electronic density of states (DOS) with 2$\Delta$(5 K)$\approx$5 meV.
The development of the peak is already indicated in the reflectivity
at 5 K by the lowering and flattening of the low-frequency part of the
spectrum (see inset of Fig.\ \ref{refl_34}).
The gap opening at low temperature strongly influences the dielectric
constant $\epsilon_1(\nu\rightarrow 0)$: Assuming a simple picture
of a semiconductor with an energy gap $\hbar\omega_g$ between the
narrow valence and conduction band, the relation $\epsilon_1(\nu=0)$$\approx$$\omega_g^{-2}$
is expected to hold.\cite{Dressel01} This explains the measured
spectrum $\epsilon_1(\nu)$ of SrNbO$_{3.41}$ for
{\bf E}$\parallel$$a$ (Fig.\ \ref{eps1}), where at 5 K high positive values
are found for $\nu$$\rightarrow$0. In contrast, at 300 K
$\epsilon_1$($\nu$$\rightarrow$0) is negative as expected for a metal.
These findings confirm the high-resolution ARPES data, which show a gap
between the leading edge midpoint of the spectrum and $E_F$ of
$\Delta$(25 K)$\approx$4 meV for $\bar\Gamma$-$\bar X$.\cite{Kuntscher02}
The gap opening as observed by optics and ARPES along the chain direction
is in agreement with the dc resistivity data along the $a$ direction, although 
there the upturn occurs at a lower temperature ($\approx$50 K) than the development
of the low-frequency peak in the optical conductivity (below 100 K), as was
already pointed out in Ref. \onlinecite{Kuntscher02}. In Fig.\ \ref{sig_34} 
we also included the dc conductivity results;\cite{Lichtenberg01} above 50 K
the dc values are obviously lower than the optical conductivity values at 
5 cm$^{-1}$. This suggests that the optical conductivity $\sigma_1$ cannot 
be described by a simple Drude-type behavior. Fitting the $\sigma_1$($\omega$)-spectra
with the Drude-Lorentz model indeed shows that a fit of reasonable quality (not shown) 
requires the introduction of two Drude peaks, e.g., with 
$\sigma_{1,1}(\omega\rightarrow$0)=2006 $\Omega^{-1}$cm$^{-1}$ and 
$\sigma_{1,2}(\omega\rightarrow$0)=2356 $\Omega^{-1}$cm$^{-1}$ at 300 K, 
which is in good agreement with the corresponding dc conductivity value. 

\subsection{SrNbO$_{3.50}$: Optics}
\label{subsectionsrnbo35}
\begin{figure}
\hspace*{5mm} \leavevmode \epsfxsize=65mm \epsfbox{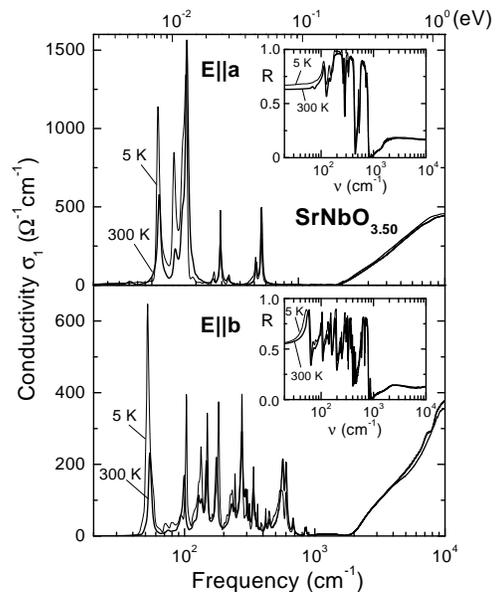}
\caption{Optical conductivity $\sigma_1$ of SrNbO$_{3.50}$ at
300 and 5 K for {\bf E}$\parallel$$a$ and {\bf E}$\parallel$$b$, respectively.
Insets: Reflectivity $R$ used for Kramers-Kronig analysis.}
\label{sr35op}
\end{figure}

The reflectivity spectra of SrNbO$_{3.50}$ (insets of Fig.\ \ref{sr35op})
are significantly different from those of SrNbO$_{3.41}$. The overall low
reflectivity for both polarization directions is in accordance with the insulating
nature of the material. The non-conducting character is also illustrated in the
optical conductivity spectra (Fig.\ \ref{sr35op}) which do not
contain electronic excitations in the low-frequency range (the onset of
interband transitions is around 2000 cm$^{-1}$ for both polarizations)
but only vibrational excitations. Perpendicular to the chains, i.e., for
{\bf E}$\parallel$$b$, a rich phonon spectrum is found.
At RT one can resolve 24 infrared active modes for {\bf E}$\parallel$$b$,
with the strongest around 54, 100, 147, 176, 275, 295, 338, 567, and
601 cm$^{-1}$, and 12 modes for {\bf E}$\parallel$$a$, the strongest
being around 112, 173, 295, and 561 cm$^{-1}$. A qualitative
assignment of the phonon modes is given in the discussion.
The mode around 54 cm$^{-1}$ for {\bf E}$\parallel$$b$ is close to the soft
mode found by Raman scattering experiments\cite{Ohi85} which is related
to the ferroelectric transition at 1615 K. It is expected to be infrared
active, and indeed was already observed in earlier reflectivity and transmission
studies.\cite{Buixaderas01}
During cooling down additional modes appear in agreement with the results of
Ref.\ \onlinecite{Buixaderas01}, and at 5 K there are altogether
24 and 51 modes resolved for {\bf E}$\parallel$$a$ and {\bf E}$\parallel$$b$,
respectively. The number of observed modes at RT and 5 K is lower than
expected from the factor group analysis; \cite{Buixaderas01} the missing
modes might be either masked by the other contributions or too weak to
be resolved.

\subsection{SrNbO$_{3.45}$: Optics}    \label{subsectionsrnb345}
\begin{figure}
\hspace*{5mm} \leavevmode \epsfxsize=65mm \epsfbox{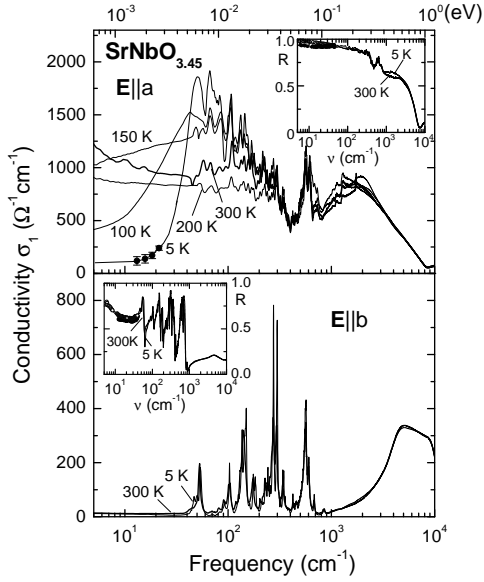}
\caption{Optical conductivity $\sigma_1$ of SrNbO$_{3.45}$
at different temperatures for {\bf E}$\parallel$$a$ and
{\bf E}$\parallel$$b$, respectively; closed circles: data points directly
determined by transmission measurements on a thin platelet.
For {\bf E}$||$$a$ the low-temperature peak in $\sigma_1$ with its
maximum around 60 cm$^{-1}$ indicates excitations across a
gap 2$\Delta$$\approx$7 meV.
Insets: Reflectivity $R$ used for Kramers-Kronig analysis;
open (closed) circles: submm reflectivity data at 300 K (5 K).}
\label{sr_345}
\end{figure}

Previous ARPES measurements\cite{Kuntscher00,Lu97} on SrNbO$_{3.45}$
revealed a quasi-1D band structure near $E_F$ similar to that of SrNbO$_{3.41}$.
Also the optical response shown in Fig.\ \ref{sr_345} clearly demonstrates
strong similarities in the electronic properties between the two compounds:
The {\bf E}$\parallel$$a$ reflectivity $R$ at RT is high and almost 1 at
low frequencies and shows a pseudo plasma edge, whereas for
{\bf E}$\parallel$$b$ the overall reflectivity is low and
no plasma edge is found.
Correspondingly, for {\bf E}$\parallel$$a$ the optical conductivity
contains a Drude-like peak due to itinerant electrons which is superimposed
by vibrational modes between 50 and 800 cm$^{-1}$, followed by a
pronounced MIR band centered around 2000 cm$^{-1}$ with a vibrational fine
structure. In contrast, for {\bf E}$\parallel$$b$ no Drude peak is
observed in $\sigma_1(\nu)$ but a large number of vibrational
excitations up to 900 cm$^{-1}$ followed by interband transitions
above $\approx$2000 cm$^{-1}$. Thus, also SrNbO$_{3.45}$ shows a
quasi-1D metallic character at RT, with a metal-like behavior along
the chains and an insulating behavior along the perpendicular direction.

The temperature dependence of the low-frequency optical properties along the 
chains agrees qualitatively with the dc resistivity,\cite{comment04} being almost 
temperature-independent between 300 K and $\approx$100 K and increases 
below 100 K with decreasing temperature.
Substantial changes occur in the low-frequency ($\leq$50 cm$^{-1}$) optical
response of SrNbO$_{3.45}$ at low temperatures ($<$100 K): 
One observes a lowering and flattening of the reflectivity
(see inset of Fig.\ \ref{sr_345}) which signals the development of a new
excitation. According to the optical conductivity (Fig.\ \ref{sr_345})
between 0 and 400 cm$^{-1}$ a redistribution of spectral weight from low
to high frequencies occurs for temperatures below 150 K, resulting in the
development of a relatively broad and asymmetric peak, which has a maximum
around 60 cm$^{-1}$ ($\approx$7 meV) and is superimposed by a large number
of phonon excitations. The peak intensity increases with decreasing
temperature. One can ascribe the development of the low-frequency peak to
the opening of an energy gap with decreasing temperature. The presence of
a small gap at low temperature is also indicated by the high values of the
corresponding dielectric constant $\epsilon_1$ at 5 K for $\nu\rightarrow$0
(Fig.\ \ref{eps1}), as was discussed for SrNbO$_{3.41}$ in section
\ref{subsectionsrnbo341}. In summary, SrNbO$_{3.45}$ resembles the electronic
properties of SrNbO$_{3.41}$ by showing both a quasi-1D metallic character
and the opening of an energy gap of only a few meV size below 150 K.

\begin{figure}
\hspace*{5mm} \leavevmode \epsfxsize=65mm \epsfbox{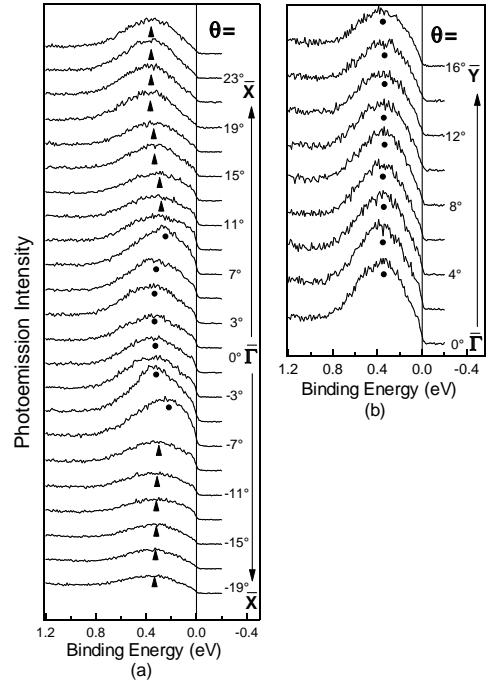}
\caption{ARPES spectra of Sr$_{0.8}$La$_{0.2}$NbO$_{3.50}$ along
(a) $\bar\Gamma$-$\bar X$ and (b) $\bar\Gamma$-$\bar Y$ at 75 K with
$\Delta$$E$=30 meV.}
\label{sr35la}
\end{figure}

\subsection{Sr$_{0.8}$La$_{0.2}$NbO$_{3.50}$: ARPES, optics}
\label{subsectionsrlanbo35}
The electronic properties of Sr$_{0.8}$La$_{0.2}$NbO$_{3.50}$ near $E_F$
were studied by ARPES along the two high-symmetry lines $\bar\Gamma$-$\bar X$
and $\bar\Gamma$-$\bar Y$; the results are depicted in Fig.\ \ref{sr35la}.
At the $\bar\Gamma$ point we observe a broad peak around 350 meV BE.
Along $\bar\Gamma$-$\bar X$ the peak slowly disperses towards $E_F$
accompanied by an intensity increase and
between $|\theta|$=7$^{\circ}$ and $|\theta|$=9$^{\circ}$ the
intensity of the peak slightly drops; at higher emission angles
a broad peak is still visible with almost constant BE ($\approx$350 meV).
Along $\bar\Gamma$-$\bar Y$ no dispersion of the broad peak is
observed. Along both directions and for all emission angles the
overall photoelectron intensity near $E_F$ is quite low.
We also studied the ARPES spectra for
Sr$_{0.7}$La$_{0.2}$Ba$_{0.1}$NbO$_{3.50}$ and
Sr$_{0.6}$La$_{0.2}$Ca$_{0.2}$NbO$_{3.50}$  (not shown), having the
same nominal carrier density as Sr$_{0.8}$La$_{0.2}$NbO$_{3.50}$
(see Table \ref{stacking}): For these two compounds the dispersion of the
broad peak along $\bar\Gamma$-$\bar X$ for $|\theta|$$\leq$7$^{\circ}$
is even less pronounced.\cite{Kuntscher00a}
The results for the (Sr,Ba,Ca)$_{0.8}$La$_{0.2}$NbO$_{3.50}$
suggest that there are two near-$E_F$ bands with
similar BEs: One band is almost dispersionless along both $\bar\Gamma$-$\bar X$
and $\bar\Gamma$-$\bar Y$; the second band has a parabolic-like
dispersion, although very weak, along $\bar\Gamma$-$\bar X$ with an
intensity drop near $E_F$ suggestive of a FS crossing, and is
dispersionless along $\bar\Gamma$-$\bar Y$. To some extent the band
dispersions of Sr$_{0.8}$La$_{0.2}$NbO$_{3.50}$ thus resemble those
of SrNbO$_{3.41}$ (see Fig.\ \ref{srnbo341_1}). However, compared
to SrNbO$_{3.41}$ the dispersion along $\bar\Gamma$-$\bar X$ is much weaker,
i.e., the quasi-1D character of the band structure is less developed.

\begin{figure}
\hspace*{5mm} \leavevmode \epsfxsize=65mm \epsfbox{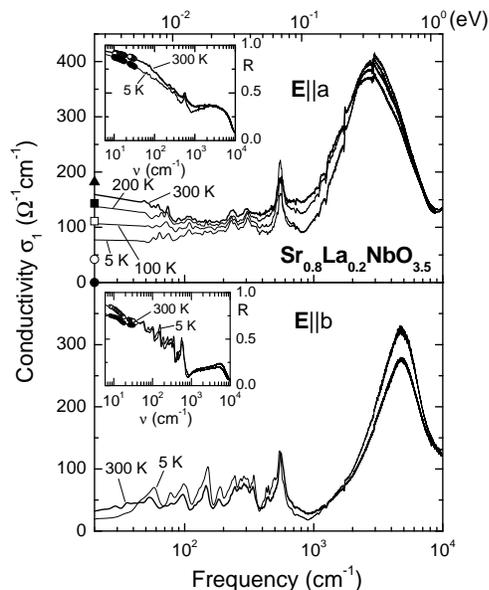}
\caption{Optical conductivity $\sigma_1$ of
Sr$_{0.8}$La$_{0.2}$NbO$_{3.50}$ at different temperatures for
{\bf E}$\parallel$$a$ and {\bf E}$\parallel$$b$, respectively.
Also shown are the dc conductivity values for the $a$ axis at 300 K 
(filled triangle), 200 K (open square), 100 K (filled square), 50 K (open circle), 
and 5 K (filled circle).
Insets: Reflectivity $R$ used for Kramers-Kronig analysis;
open (closed) circles: submm reflectivity data at 300 K (5 K).}
\label{srla35op}
\end{figure}

Further insight was gained by temperature-dependent reflectivity measurements
for {\bf E}$\parallel$$a,b$ (see Fig.\ \ref{srla35op}).
For the polarization {\bf E}$\parallel$$b$ the overall low
reflectivity and the absence of a Drude component in $\sigma_1(\nu)$
demonstrate the insulating character of the material perpendicular
to the chains. For {\bf E}$\parallel$$a$ the
reflectivity at RT is metal-like with relatively high values at low
frequencies, decreases for higher frequencies, and finally
drops rapidly above 6000 cm$^{-1}$. The corresponding optical conductivity
shows a very weak Drude-like contribution and a pronounced MIR absorption 
peak around 3000 cm$^{-1}$. Between 300 K and 50 K the low-frequency optical 
conductivity is in reasonable agreement with the dc conductivity,\cite{Lichtenberg01}
but at 5 K it is lower than the dc value. 
The overall optical response of Sr$_{0.8}$La$_{0.2}$NbO$_{3.50}$
is similar to that of SrNbO$_{3.41}$ (see Fig.\ \ref{sig_34})
and SrNbO$_{3.45}$ (see Fig.\ \ref{sr_345}), but there are important
differences: The overall reflectivity of Sr$_{0.8}$La$_{0.2}$NbO$_{3.50}$
along the chains is lower in the far-infrared range and the plasma
edge is hardly developed; both findings indicate only weak metallic
character. Thus, the optical conductivity of Sr$_{0.8}$La$_{0.2}$NbO$_{3.50}$ 
for {\bf E}$\parallel$$a$ contains a much weaker Drude-like term. 
Importantly, with decreasing temperature the development of a
low-frequency peak is not observed, in contrast to what is found
for SrNbO$_{3.41}$ and SrNbO$_{3.45}$ (see Figs.\ \ref{sig_34}
and \ref{sr_345}).

\section{Discussion}
\label{sectiondiscussion}
\begin{figure}
\hspace*{5mm} \leavevmode \epsfxsize=65mm \epsfbox{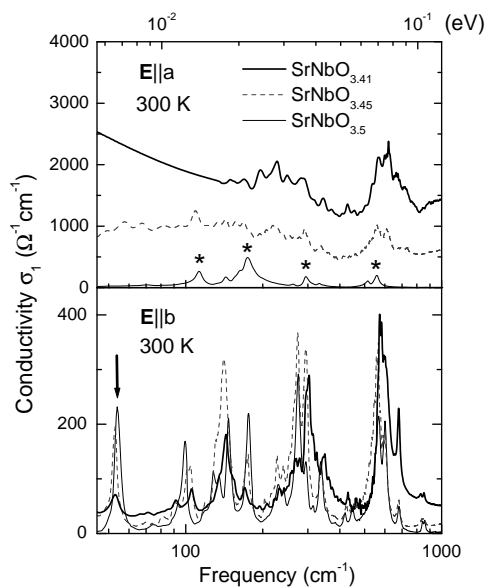}
\caption{Optical conductivity $\sigma_1$ of SrNbO$_{3.50}$,
SrNbO$_{3.45}$, and SrNbO$_{3.41}$ at RT for {\bf E}$\parallel$$a$ and
{\bf E}$\parallel$$b$, respectively. The asterisks mark the phonon lines
of SrNbO$_{3.50}$ for {\bf E}$\parallel$$a$ which survive in the
conducting compounds. The arrow indicates the ferroelectric soft
mode of SrNbO$_{3.50}$.}
\label{phonon1}
\end{figure}

According to the experimental results, the electronic
properties of the SrNbO$_{3.50-x}$ compounds strongly depend
on the oxygen content, which not only determines the crystal structure,
but also the number of conduction electrons in the material (see Table \ref{stacking}).
To illustrate these differences we plot in Fig.\ \ref{phonon1} the
optical conductivity at RT for the three compositions SrNbO$_{3.41}$,
SrNbO$_{3.45}$, and SrNbO$_{3.50}$ for {\bf E}$\parallel$$a,b$.
A contribution of itinerant electrons appears in the
{\bf E}$\parallel$$a$ spectra: A Drude peak is observed for SrNbO$_{3.45}$,
which is even more pronounced for SrNbO$_{3.41}$ in full agreement
with the higher density of conduction electrons for the latter compound,
while the absence of a Drude peak for SrNbO$_{3.50}$ accounts for its
insulating character.

\begin{figure}
\hspace*{5mm} \leavevmode \epsfxsize=65mm \epsfbox{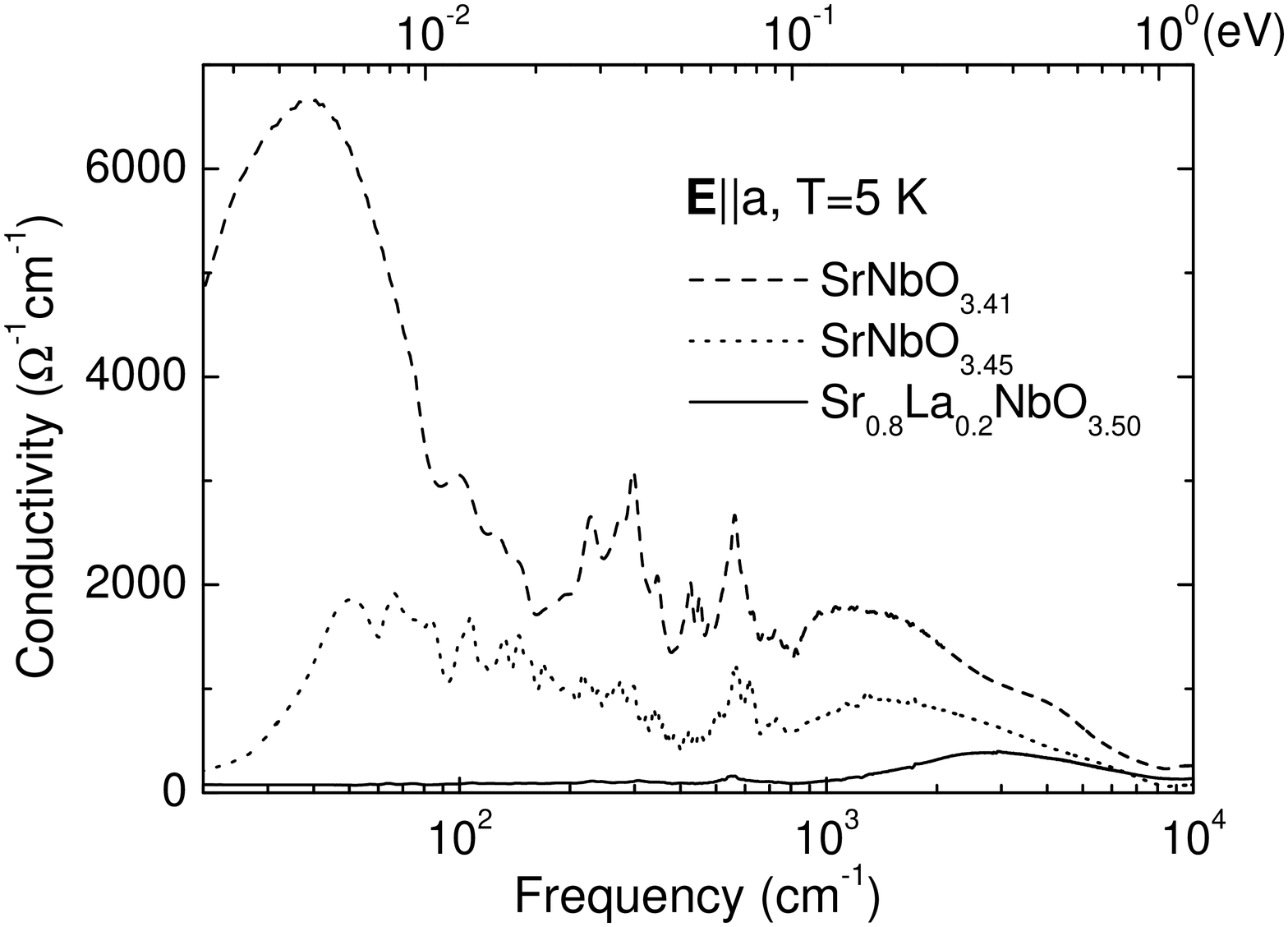}
\caption{Optical conductivity $\sigma_1$ of SrNbO$_{3.41}$,
SrNbO$_{3.45}$, and Sr$_{0.8}$La$_{0.2}$NbO$_{3.50}$ at 5 K
for {\bf E}$\parallel$$a$.}
\label{signb}
\end{figure}

Despite their layered crystal structure, SrNbO$_{3.45}$ and SrNbO$_{3.41}$
show a pronounced quasi-1D metallic character which can be related to
the chain-like arrangement of NbO$_6$ octahedra along the $a$ axis.
A more detailed picture is provided by band structure calculations
in the local density approximation (LDA) for SrNbO$_{3.41}$:\cite{Kuntscher00,Winter00}
The predominant contribution to the DOS near $E_F$ is attributed
to the Nb site in the middle of the slab, while the other two\cite{comment05}
Nb sites contribute considerably less (see Fig.\  6(a) in Ref.\ \onlinecite{Winter00}).
The corresponding NbO$_6$ octahedra in the middle of the slabs (marked by thick
lines in Fig.\ \ref{crystal0}) are the least distorted ones, with the Nb atoms
located right in the center of the octahedra; for all other octahedra
the Nb atoms are considerably shifted away from the central octahedral
position. The quasi-1D metallic character is thus related to the central
chains of least distorted octahedra in the middle of the slabs. The influence
of the central octahedral chains on the electronic characteristics is
illustrated by the experimental results of Sr$_{0.8}$La$_{0.2}$NbO$_{3.50}$
compared to those of SrNbO$_{3.41}$:
Both compounds nominally contain the same density of Nb$4d$ conduction
electrons (see Table \ref{stacking}). However, in Sr$_{0.8}$La$_{0.2}$NbO$_{3.50}$
the slabs are only {\it four} octahedra wide, i.e., the central chains are
missing. According to the resistivity\cite{Lichtenberg01} and optical results
presented here, this compound shows a much weaker metallic behavior compared
to SrNbO$_{3.41}$, and the quasi-1D character of its band
structure, as observed by ARPES, has almost vanished.
This is in agreement with the absence of a gap opening for
Sr$_{0.8}$La$_{0.2}$NbO$_{3.50}$ along the chains (see Fig.\ \ref{signb}):
The intensity of the weak Drude peak observed in the RT optical conductivity
for {\bf E}$\parallel$$a$ has decreased at 5 K, but a low-frequency peak has
not developed, in contrast to what is found for SrNbO$_{3.41}$ and SrNbO$_{3.45}$.
All these experimental findings show that the quasi-1D metallic
character and the related gap opening observed for SrNbO$_{3.41}$ and
SrNbO$_{3.45}$ are indeed linked to the central octahedral chains.

\begin{figure}
\hspace*{5mm} \leavevmode \epsfxsize=65mm \epsfbox{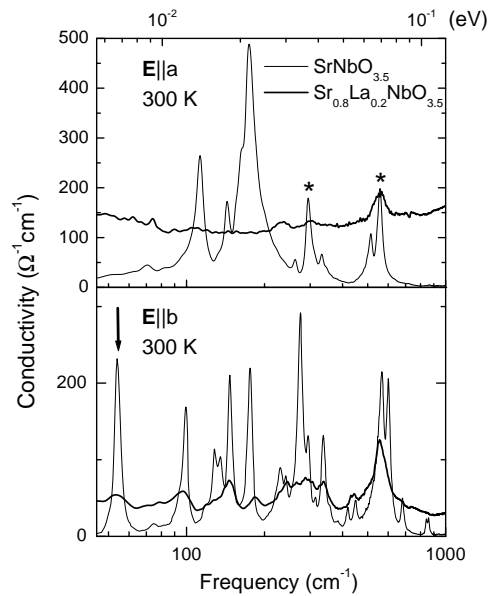}
\caption{Optical conductivity $\sigma_1$ of SrNbO$_{3.50}$
and Sr$_{0.8}$La$_{0.2}$NbO$_{3.50}$ at RT for {\bf E}$\parallel$$a$ and
{\bf E}$\parallel$$b$, respectively. The asterisks mark the phonon lines
of SrNbO$_{3.50}$ for {\bf E}$\parallel$$a$ which survive in the
conducting compound. The arrow indicates the ferroelectric soft mode of
SrNbO$_{3.50}$.}
\label{phonon2}
\end{figure}

The compounds SrNbO$_{3.50}$, SrNbO$_{3.45}$, and SrNbO$_{3.41}$ have a complex
crystal structure and thus exhibit rich phonon spectra for both polarization
directions (see Fig.\ \ref{phonon1}). For {\bf E}$\parallel$$a$
the main groups of phonon modes observed for SrNbO$_{3.50}$, which are 
centered around 110, 170, 300, and 540 cm$^{-1}$ and marked by asterisks 
in Fig.\ \ref{phonon1}, survive in the conducting members, with only
small frequency shifts and a masking due to the contribution of itinerant
electrons. The similarities in the vibrational properties are more obvious
for the polarization direction perpendicular to the chains, {\bf E}$\parallel$$b$,
since no Drude contribution is present: The main groups of phonon lines,
centered around 53, 100, 150, 290, and 600 cm$^{-1}$, are present in all three
compounds. A qualitative assignment to the lattice dynamics can be achieved by
a comparison with other niobium oxides with a similar network of corner-sharing
NbO$_6$ octahedra:
The high-frequency infrared-active modes of KNbO$_3$ (NaNbO$_3$) were assigned
to the internal modes of NbO$_6$ octahedra, namely the 660 (675) and
550 (510) cm$^{-1}$ modes to the stretching modes and the 375 (375) cm$^{-1}$ mode
to the bending mode.\cite{Last57} This is in agreement with later Raman
and infrared measurements on LiNbO$_3$ and Ba$_2$NaNb$_5$O$_{15}$.\cite{Ross70}
The lower-frequency lines ($\nu$$<$200 cm$^{-1}$) were assigned to
translational modes of the cations.\cite{Ross70}
The influence of La doping on the optical properties is illustrated in
Fig.\ \ref{phonon2} where we plot the conductivity spectra of SrNbO$_{3.50}$
and Sr$_{0.8}$La$_{0.2}$NbO$_{3.50}$.
For {\bf E}$\parallel$$a$ an electronic background related to itinerant
carriers in the Sr$_{0.8}$La$_{0.2}$NbO$_{3.50}$ spectrum
is visible. Two of the main groups of phonon modes of
SrNbO$_{3.50}$ (marked by asterisks) are also present in the (weakly) conducting
material, but they are broader and strongly damped.
Interestingly, the relatively strong modes observed in SrNbO$_{3.50}$
around 110 and 170 cm$^{-1}$ appear to be completely suppressed in
Sr$_{0.8}$La$_{0.2}$NbO$_{3.50}$, indicating a strong damping
of the cation translational modes by partial substitution of Sr by La.
Along the $b$ direction the similarities in the vibrational
properties are also obvious.
In particular, the mode around 54 cm$^{-1}$ for
{\bf E}$\parallel$$b$, indicated by an arrow in Fig.\ \ref{phonon1} and
\ref{phonon2} is present in all four compounds discussed here. For
SrNbO$_{3.50}$ this mode was assigned to the soft mode of the ferroelectric
phase transition at 1615 K, based on Raman scattering experiments.\cite{Ohi85}
It was also observed in recent optical measurements and related to
NbO$_6$ octahedra tiltings.\cite{Buixaderas01}
In agreement with Ref.\ \onlinecite{Buixaderas01}, this soft mode is
temperature independent between 300 and 100 K and shifts to higher
frequencies below 100 K. The precise temperature dependence is, however,
difficult to determine due to the overlap with another mode which appears
around 50 cm$^{-1}$ at $\approx$100 K during cooling down.

Finally, we want to discuss the puzzling electronic properties of
the conducting members of the present niobate series at low
temperature, in particular the opening of an energy gap at $E_F$
found for SrNbO$_{3.41}$ and SrNbO$_{3.45}$. Several aspects make
an explanation of the gap opening in terms of a Peierls-type
scenario most likely: (i) The FS nesting\cite{Kuntscher00} could
drive the lattice towards a Peierls distortion; (ii) the shadow
band observed for SrNbO$_{3.41}$ in ARPES\cite{Kuntscher02} would
be consistent with a superstructure with wave vector
$q$=2$k_F$;\cite{Voit00} (iii) the appearance of phonon lines with
decreasing temperature suggests a structural phase transition such
as a Peierls instability. 
However, there remain some inconsistencies within a Peierls scenario: 
In contrast to what is expected for CDW formation, no sharp anomaly 
is found in dc resistivity; neither do thermal expansion 
measurements\cite{Meingast00} show indications of a structural phase 
transition. Furthermore, an estimate of the mean-field transition
temperature based on the observed gap size is not compatible with 
the temperature development of dc resistivity and optical 
conductivity.\cite{Kuntscher02}

Further open issues are the transport mechanism in SrNbO$_{3.41}$ and 
SrNbO$_{3.45}$ along the chain direction and the closely related 
temperature dependence of the dc resistivity $\rho_a$, showing different
regimes. The strong upturn of $\rho_a$ below 50 K with cooling down can 
be explained by the opening of a gap, in agreement with ARPES and optics. 
The temperature dependence above 50 K may be due to contributions
of quasiparticles with different
characteristics. While the metallic temperature dependence between
50 K and 130 K suggests a band-like motion of electrons, above 
130 K the hopping of more localized charge carriers appears to dominate. 
A possible mechanism for the localization/trapping of charge carriers
could be strong electron-phonon coupling, which might also explain the 
observation of a pronounced MIR absorption band in the optical 
conductivity along the chain direction (see Figs.\ \ref{sig_34} and 
\ref{sr_345}). Similar absorption bands in related quasi-1D metallic 
compounds were recently interpreted in terms of excitations of polaronic
quasiparticles.\cite{Presura03,Kuntscher03} Further spectroscopic 
information, e.g., from temperature-dependent ARPES measurements, is needed
to clarify a possible polaron formation in the studied niobates.

In conclusion, we determined the electronic and vibrational properties of the
perovskite-related compounds Sr$_{1-y}$La$_y$NbO$_{3.5-x}$ and related
the results to their crystal structure. A possible transport mechanism
of SrNbO$_{3.41}$ and SrNbO$_{3.45}$ was discussed and the importance 
of electron-phonon coupling in these systems suggested, possibly leading
to polaron formation. To explain the low-temperature behavior a Peierls 
picture appears to be the most appropriate, but is insufficient to describe
all the findings. An important aspect might be the close structural 
affinity of the quasi-1D metals SrNbO$_{3.41}$ and SrNbO$_{3.45}$ to the 
ferroelectric SrNbO$_{3.50}$: SrNbO$_{3.41}$
(and the ($n$=5)-slabs in SrNbO$_{3.45}$) contains the same structural
building blocks as SrNbO$_{3.50}$, but with additional chains of basically
{\it undistorted} NbO$_6$ octahedra inserted in the center of the slabs.
Ferroelectric SrNbO$_{3.50}$ shows relatively high values of the static dielectric
constant $\epsilon_1$ at RT ($\epsilon_{1,a}$=75, $\epsilon_{1b}$=43, and
$\epsilon_{1,c}$=46)\cite{Nanamatsu75} indicating highly polarizable
structural units. A further important result is the observation
of the phonon mode at 54 cm$^{-1}$ in the {\bf E}$\parallel$$b$ conductivity
spectra of all present niobate compounds (see Fig.\ \ref{phonon1}
and \ref{phonon2}), which for SrNbO$_{3.50}$ can be assigned to the
ferroelectric soft mode. Since the soft modes in ferroelectrics in general
lead to electrical polarization, the presence of the 54 cm$^{-1}$ mode
in SrNbO$_{3.41}$ and SrNbO$_{3.45}$ indicates a polarizability of their
lattices. The highly conducting channels of octahedra in the middle of
the slabs, which are responsible for the 1D metallic transport, can thus
be viewed as embedded in a highly polarizable medium, and a certain
influence of this medium on the electronic properties of the chains is
conceivable. This picture is, however, speculative since to our knowledge 
a 1D model taking into account such a polarizable medium surrounding 
the conducting entities does not exist.

\section{Summary}
By angle-resolved photoemission spectroscopy and polarization-dependent
reflectivity measurements in the infrared we studied the electronic
and vibrational properties of several members of the series Sr$_{1-y}$La$_y$NbO$_{3.50-x}$
(0$\leq$$x$$\leq$0.1) along and perpendicular to the chains of NbO$_6$
octahedra. SrNbO$_{3.50}$ is a ferroelectric insulator with a rich
phonon spectrum for both polarization directions {\bf E}$\parallel$$a,b$.
At room temperature the compounds SrNbO$_{3.41}$ and SrNbO$_{3.45}$ exhibit
a metallic behavior with a strongly dispersing band
along the chains ({\bf E}$\parallel$$a$) and an insulating, nondispersive
character in the perpendicular direction ({\bf E}$\parallel$$b$); they are
thus quasi-1D metals.
Despite the similar carrier density, for Sr$_{0.8}$La$_{0.2}$NbO$_{3.50}$
the quasi-1D metallic character is much less developed.
Based on a comparison between the experimental results of
SrNbO$_{3.41}$, SrNbO$_{3.45}$, and
Sr$_{0.8}$La$_{0.2}$NbO$_{3.50}$ the quasi-1D metallic character observed
in this series can be related to the central chains of least
distorted octahedra located in the middle of the slabs. This
is in full agreement with the results from LDA band structure calculations.
We propose a possible transport mechanism for the compounds SrNbO$_{3.41}$ 
and SrNbO$_{3.45}$, with a band-like contribution of mobile carriers and 
a contribution due to the hopping motion of more localized charges 
dominating above 130 K. 
At low temperature the opening of an extremely small energy gap at $E_F$ 
is observed for SrNbO$_{3.41}$ and SrNbO$_{3.45}$
along the chain direction, which might be related to a 
Peierls-type instability, with electron-phonon coupling as the driving
mechanism. However, not all findings can be explained within a Peierls 
picture, and we speculate that the highly polarizable structural units 
surrounding the conducting channels might have an appreciable influence.

\subsection*{Acknowledgements}
We are grateful to L. Perfetti, C. Rojas, and I. Vobornik
for valuable technical help. We thank H. Winter for fruitful discussions.
Financial support by the DAAD, the BMBF (project No.\
13N6918/1), and the Deutsche Forschungsgemeinschaft is gratefully acknowledged.


\begin{references}
\item[$^*$] Present address: General Physics Institute,
Russian Academy of Sciences, Vavilov 38, Moscow 117942, Russia.

\bibitem{Kuntscher00}
C. A. Kuntscher, S. Gerhold, N. N\"ucker, T. R. Cummins, D.-H. Lu, S. Schuppler,
C. S. Gopinath, F. Lichtenberg, J. Mannhart, K.-P. Bohnen,
Phys.\ Rev.\ B {\bf 61}, 1876 (2000).

\bibitem{Kuntscher02}
C. A. Kuntscher, S. Schuppler, P. Haas, B. Gorshunov, M. Dressel,
M. Grioni, F. Lichtenberg, A. Herrnberger, F. Mayr, and  J. Mannhart,
Phys.\ Rev.\ Lett.\ {\bf 89}, 236403 (2002).

\bibitem{Lichtenberg01}
F. Lichtenberg, A. Herrnberger, K. Wiedenmann, and J. Mannhart,
Prog. Solid State Chem {\bf 29}, 1 (2001).

\bibitem{Abrahams98}
S. C. Abrahams, H. W. Schmalle, T. Williams, A. Reller, F. Lichtenberg, D. Widmer,
J. G. Bednorz, R. Spreiter, Ch.\ Bosshard, and P. G\"unter, Acta Cryst.\ B {\bf 54},
399 (1998).

\bibitem{Williams93}
T. Williams, F. Lichtenberg, D. Widmer, J. G. Bednorz, and A. Reller,
J. Solid State Chem.\ {\bf 103}, 375 (1993).

\bibitem{Bednorz97}
J. G. Bednorz, K. H. Wachtmann, R. Broom, and D. Ariosa, in
{\it High-T$_c$ superconductivity 1996: Ten Years after the discovery},
edited by E. Kaldis, E. Liarokapis, and K. A. M\"uller
(Kluwer, Dordrecht, 1997).

\bibitem{Nanamatsu71}
S. Nanamatsu, M. Kimura, K. Doi und M. Takahashi, J. Phys.\ Soc.\ Jpn.\
\textbf{30}, 300 (1971);

\bibitem{Nanamatsu75}
S. Nanamatsu, M. Kimura, and T. Kawamura, J. Phys.\ Soc.\ Jpn.\ {\bf 38}, 817 (1975).

\bibitem{Ohi79}
K. Ohi, M. Kimura, H. Ishida, and H. Kakinuma, J. Phys.\ Soc.\ Jpn.\
\textbf{46}, 1387 (1979).

\bibitem{Weber01}
J.-E. Weber, C. Kegler, N. B\"uttgen, H.-A. Krug von Nidda, A. Loidl,
and F. Lichtenberg, Phys.\ Rev.\ B {\bf 64}, 235414 (2001).

\bibitem{Bobnar02}
V. Bobnar, P. Lunkenheimer, J. Hemberger, A. Loidl, F. Lichtenberg, and
J. Mannhart, Phys.\ Rev.\ B {\bf 65}, 155115 (2002).

\bibitem{Kozlov98}
G. V. Kozlov and A. A. Volkov, in {\it Millimeter and Submillimeter Wave
Spectroscopy of Solids}, ed.\ G. Gr\"uner (Springer, Berlin, 1998).

\bibitem{Dressel01}
M. Dressel and G. Gr\"uner, {\it Electrodynamics in solids: Optical properties of
electrons in matter} (Cambridge University Press, Cambridge, 2002).

\bibitem{Kuntscher00a}
C. A. Kuntscher, PhD thesis, Universit\"at Karlsruhe
(Cuvillier Verlag, G\"ottingen, 2000).

\bibitem{Pintschovius}
L. Pintschovius, unpublished.

\bibitem{Ohi85}
K. Ohi and S. Kojima, Japan.\ J. Appl.\ Phys.\ {\bf 24}, 817 (1985).

\bibitem{Buixaderas01}
E. Buixaderas, S. Kamba, and J. Petzelt, J. Phys.: Condens.\ Matter
{\bf 13}, 2823 (2001).

\bibitem{Lu97}
D. H. Lu, C. S. Gopinath, M. Schmidt, T. R. Cummins, N. N\"ucker, S. Schuppler,
and F. Lichtenberg, Physica C {\bf 282-287}, 995 (1997).

\bibitem{comment04}
No dc resistivity data are available for SrNbO$_{3.45}$, but it is expected
to be qualitatively similar to those of Sr$_{0.96}$Ba$_{0.04}$NbO$_{3.45}$ 
which has the same crystal structure\cite{Lichtenberg01} and, within an ionic
picture, the same electron doping as SrNbO$_{3.45}$.

\bibitem{Winter00}
H. Winter, S. Schuppler, and C. A. Kuntscher,
J. Phys.\ Cond.\ Matt.\ {\bf 12}, 1735 (2000).

\bibitem{comment05}
There are three inequivalent Nb sites per unit cell in SrNbO$_{3.41}$,
as was illustrated in Ref.\ \onlinecite{Winter00}.

\bibitem{Last57}
J. T. Last, Phys.\ Rev.\ {\bf 105}, 1740 (1957);

\bibitem{Ross70}
S. D. Ross, J. Phys.\ C {\bf 3}, 1785 (1970).

\bibitem{Voit00}
J. Voit, L. Perfetti, F. Zwick, H. Berger, G. Margaritondo, G. Gr\"uner,
H. H\"ochst, and M. Grioni, Science {\bf 290}, 501 (2000).

\bibitem{Meingast00}
P. Nagel, V. Pasler, and C. Meingast, unpublished.

\bibitem{Presura03}
C. Presura, M. Popinciuc, P. H. M. van Loosdrecht, D. van der Marel, M. Mostovoy,
T. Yamauchi, and Y. Ueda,
Phys.\ Rev.\ Lett.\ {\bf 90}, 026402 (2003).

\bibitem{Kuntscher03}
C. A. Kuntscher, D. van der Marel, M. Dressel, F. Lichtenberg, and J. Mannhart,
Phys.\ Rev.\ B {\bf 67}, 035105 (2003).


\end{references}
\end{document}